\documentclass[aps,twocolumn,prd,showpacs,nofootinbib]{revtex4}
\usepackage{amsmath}
\usepackage{graphicx}
\usepackage{dcolumn}
\usepackage{bm}
\usepackage{amssymb}
\usepackage{latexsym}

\def\be{\begin{equation}}
\def\ee{\end{equation}}
\def\ba{\begin{eqnarray}}
\def\ea{\end{eqnarray}}

\newcommand{\bd}{\ensuremath{\overline{\textnormal{D3}}}}

\begin{document}

\title{``Phantom" Inflation in Warped Compactification}

\author{Yun-Song Piao}

\affiliation{College of Physical Sciences, Graduate School of
Chinese Academy of Sciences, Beijing 100049, China}

\begin{abstract}

In this paper, in a class of warped compactifications with the
brane/flux annihilation, we find that the inflation may be driven
by a flat direction identified as that along the number p of
\bd-branes located at the tip of the Klebanov-Strassler throat.
The spectrum of adiabatic perturbation generated during inflation
is nearly scale invariant, which may be obtained by using the
results shown in the phantom inflation, since in a four-dimension
effective description the evolution of energy density along the
$p$ direction is slowly increasing.


\end{abstract}


\maketitle

The results of recent observations are consistent with an
adiabatic and nearly scale invariant spectrum of primordial
perturbations, as predicted by the simplest models of inflation.
The inflation stage is supposed to have taken place at the earlier
moments of the universe \cite{Guth, LAS, S80}, which
superluminally stretched a tiny patch to become our observable
universe today. During the inflation the quantum fluctuations in
the horizon will be able to leave the horizon and become the
primordial perturbations responsible for the structure formation
of observable universe
\cite{BST83, MC}. Recently, how embedding the inflation scenario
into string theory has received increasing attentions, e.g. see
Ref. \cite{Tye06, C, Linde07, Bur, MS} for reviews.


The inflation can be generally regarded as an accelerated or
superaccelerated stage, and so may defined as an epoch when the
comoving Hubble length decreases. This also may occurs samely
during a null energy condition violating expansion which
corresponds to e.g. the phantom inflation \cite{PZ, PZ1, GJ}, see
also Ref. \cite{DHC} for comments. The phantom inflation is only a
phenomenological description for such a superaccelerated stage, it
dose not equal to the existence of a phantom. Here we will search
for a realistic implement of the phantom inflation in possible
high energy physical theory, e.g. the string theory. Recently, it
has been shown in Ref. \cite{KPV} that in a class of string
compactification with the warped metric, when there are $p\ll
{\cal M}$ \bd-branes inside the KS throat, which is a deformed
conifold with ${\cal M}$ units of RR 3-form flux around the
internal $S^3$, where $p$ is the number of \bd-branes, a
metastable and nonsupersymmetric bound state with positive vacuum
energy density will form. This false vacuum may be responsible for
the inflation, which has awaken a string version of old inflation
\cite{PRZ}. When $p$ is beyond some critical value $p_c$, the
false vacuum will disappear and the system will roll down to the
supersymmetric stable vacuum, which signs the end of inflation. In
the language of NS5-branes, what happens is that the \bd-brane
``blows up" into a NS5-brane wrapping a $S^2$ in the internal
$S^3$. The resulting state for $p\ll {\cal M}$ is a metastable
state which breaks supersymmetry. When $p$ is close to ${\cal M}$,
the radius of $S^2$ will approach and surmount the radius of the
equator of $S^3$, which leads that the NS5-brane becomes
perturbatively unstable, and will directly rolls down to a
supersymmetric vacuum.
In this note, we find that under some conditions there may be a
flat direction along the number $p$ of \bd-branes in this
scenario, which may lead to a nearly scale invariant spectrum of
adiabatic perturbation. The spectrum can be obtained by using that
similar to the phantom inflation \cite{PZ, PZ1}, since in the four
dimension effective description the climbing up of energy density
along $p$ direction corresponds to an evolution with the null
energy condition violation.

We begin with a 10 dimension CY manifold with the warped KS throat
\cite{KS}. The metric of warped throat may be taken as \be ds^2=
{1\over \sqrt{f(r)}}ds^2_{(4)}+\sqrt{f(r)}(dr^2+r^2 ds_{(5)}^2)
\ee for $r<r_*$, where $ds^2_{(5)}$ is the angular part of the
internal metric and $r$ is the proper distance to the tip of the
throat. When $r>r_*$, this metric can be glued to the metric of
the bulk of the compact space, usually taken to be a CY manifold.
The warp factor $f(r)$ has a minimal value at $r_0$, which is
determined by $\beta\equiv {r_0\over R}\sim e^{-{2\pi {\cal
K}\over 3g_s {\cal M}}}$, and when $r_0< r <r_{*}$, $f(r)$ may be
approximately regarded as $f(r)= ({R\over r})^4$, where $R^4
={27\pi\over 4} g_s N\alpha^{\prime 2}$, $N$ is equal to the
product of the fluxes ${\cal M}$ and ${\cal K}$ for the RR and
NSNS 3-forms, respectively, and $g_s$ is the string coupling and
$\alpha^\prime$ is set by the string scale.

It has been shown in Ref. \cite{KPV} that when putting $p\ll {\cal
M}$ \bd-branes at the tip of the KS throat,
the system will be relaxed to a nonsupersymmetric NS5-brane
``giant graviton'' configuration, in which the NS5-brane warps a
$S^2$ in $S^3$, and carries $p$ unites flux for 2-form, which
induces the \bd-charge. The $S^2$ is inclined to expand as a
spherical shell in $S^3$. This process may be parameterized by an
angle $0 \leqslant \psi \leqslant \pi $, where $\psi =0$
corresponds to the north pole of $S^3$ and $\psi =\pi$ is the
south pole. The angular position indeed appears as a scalar in the
world volume action, which describes the motion of the NS5-brane
across the $S^3$. The total effective potential can be given by
\be V_{\rm eff}(\psi)= {\cal M}\beta^4 T_3\left(\sqrt{{b_0^2
\sin^4{\psi}\over \pi^2}+{\tilde V}^2(\psi)}+{\tilde
V}(\psi)\right) \label{vpsi}\ee with $b_0\simeq 0.9$, where
${\tilde V}(\psi) ={p\over {\cal M}}-{\psi-\sin{(2\psi)/2}\over
\pi}$ and $T_3$ is the \bd-brane tension. This potential is
plotted in Fig.1 with respect to $\psi$ and $p/{\cal M}$. The
increase of potential energy along $p$ direction is actually
discrete, since $p$ is the positive integer. However, since ${\cal
M}$ is quite large, thus $p/{\cal M}$ may be approximately
regarded as continual one.

In the regime with $p/{\cal M} < 0.08$, the system sits on a
metastable state corresponding to a static NS5-brane wrapping a
$S^2$ in $S^3$, which is classically stable false vacuum. It only
may decay to a supersymmetric state by quantum tunneling, however,
as has been shown that this probability is exponentially
suppressed \cite{KPV}. In Fig.1, it can be clearly seen that this
metastable bound state corresponds to $\psi = 0$, thus the energy
density of this metastable vacuum is $ V_{\rm
eff}(\psi=0)=2p\beta^4 T_3$ given by Eq.(\ref{vpsi}). It is this
vacuum energy that drives inflation. Thus the Hubble expansion is
given by \be h^2= {2p\beta^4 T_3 \over 3}, \label{h}\ee where
$8\pi/m_p^2= 1$ is set. While the true minimum is at $\psi = \pi$,
in which the potential energy is $0$.

In the regime $p/{\cal M} \gtrsim 0.08$, the false vacuum
disappears, which means that the nonsupersymmetric configuration
of $p$ \bd-branes becomes classically unstable and will relaxes to
the supersymmetric minimum by a classical rolling, which denotes
that the \bd-branes cluster to form the maximal NS5-branes, and
then rolls down towards the bottom of the potential\footnote{It
should be mentioned that this classic rolling may also support a
slow roll inflation under some conditions, which has been studied
in Ref. \cite{DKV}. }, at the north pole $\psi =\pi$, in which the
potential energy is $0$. Thus in this case, the inflation will be
expected to end. The result of this process is ${\cal M}-p$
D3-branes instead of the original $p$ \bd-branes appearing in the
tip of KS throat, while the 3-form flux ${\cal K}$ is changed to
${\cal K}-1$, which is referred to as brane/flux annihilation in
Ref. \cite{KPV}.




The point here is that the inflation occur only when $p/{\cal M} <
0.08$, and during this period some \bd-branes may be expected to
continually enter into the throat which will continually increase
the value of $p$, and when $p/{\cal M} \gtrsim 0.08$, the
metastable minimum of the potential along $\psi$ will disappear
and the potential becomes a monotonic decreasing function of
$\psi$, and the exit of inflation is induced by a subsequent
rolling along $\psi$, see the dashed line in Fig.1.
Thus this inflation model is actually quite similar to the hybrid
inflation in Ref. \cite{L94}, in which the inflation is ended by a
water fall reheating when the inflaton reaches some critical
value. In this analogy, $\psi$ is actually identified as a water
fall field in the hybrid inflation\footnote{In Ref. \cite{PRZ},
$\psi$ was regarded as inflaton, which is different from the point
here. The curvature perturbation in their model is given by the
curvaton mechanism \cite{LW, LM}, since the curvature of the
potential of $\psi$ around the metastable minimum is much larger
than the Hubble rate during inflation, which implies that the
curvature perturbation associated to the quantum fluctuations of
$\psi$ is heavily blue tilt and thus suppressed on large scale.}.
However, the distinguish is that here the energy density driving
the inflation is increased before $p$ approaches the critical
value, which means that in the four dimension effective
description the null energy condition is actually
violated\footnote{It is interesting to note that the super
inflation phase with the null energy condition violation also
appears in loop quantum gravity, which have been studied in
Refs.\cite{B, MN}, and also \cite{ZL}, or other gravity theories
with high order corrections.
} , however, in the viewpoint of high dimension the evolution of
\bd-branes dose not actually violate the null energy condition.
The change of energy density along $p$ direction may be depicted
by the parameter $\epsilon$, which has been extensively used and
is defined as $-{\dot h}/h^2$. It can be rewritten as $\epsilon
\simeq -{1\over h \Delta t}({\Delta h\over h})$. In unit of Hubble
time $ 1/h$, we have $|\epsilon | \simeq {\Delta h\over h}\simeq
{\Delta p\over (2p)}$, where Eq.(\ref{h}) has been used and
$\Delta p$ is the change of $p$ in unit of Hubble time.
The inflation requires $|\epsilon |\ll 1$, which means that in
unit of Hubble time ${\Delta p/ p} \ll 1$. This may be assured by
a large $p$ and a small change of $p$ at the time of about 60
efolding number and after it. For instance, if in unit of Hubble
time the number $\Delta p$ of \bd-branes pulled into the throat is
about $\Delta p\sim 10$, we may take ${\cal M}\sim 10^4$ and
$p\simeq 500$, and thus have $|\epsilon | \simeq {\Delta p/ (2p)}
\simeq 0.01$, while $p/{\cal M}\simeq 0.05$ is also in required
regime.

\begin{figure}[t]
\begin{center}
\includegraphics[width=9cm]{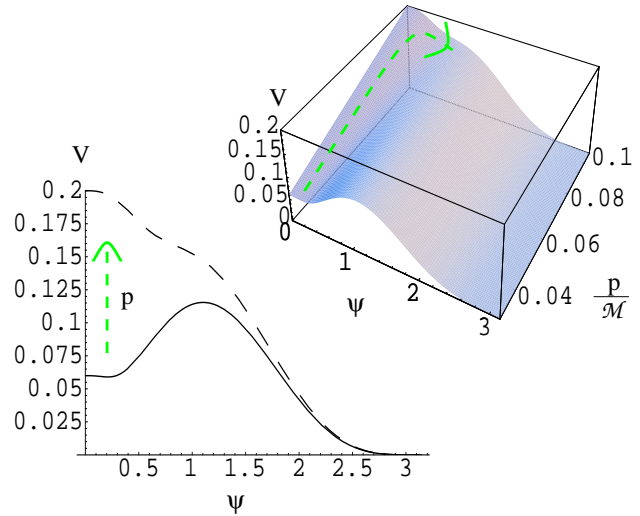}
\caption{The figure of the potential (\ref{vpsi}) up to an overall
scale ${\cal M}\beta^4 T_3$. In the left lower panel, the solid
line is the effective potential with $p/{\cal M}=0.03$. When
\bd-branes are pulled into the throat continuously, the metastable
minimum will rise inch by inch. The dashed line is the effective
potential with $p/{\cal M}=0.1$. The right up panel is the figure
of the effective potential with respect to $\psi$ and $p/{\cal
M}$. The dashed line denotes the evolution of energy density along
$p$ direction during inflation and the rolling down along $\psi$
direction after inflation. The total process is quite similar to
the hybrid inflation \cite{L94}. However, the distinguish is that
here the energy density driving the inflation is increasing before
$p$ approaches the critical value.  }
\end{center}
\end{figure}

Those \bd-branes walking into the throat should be originally in
the compactification bulk at $r
> r_*$. Their initial energy at $r = r_*$ is obtained by
multiplying the brane tension with the redshift factor $r_*/R$. It
is required in an effective theory that the \bd-brane should be a
small perturbation of the system and the vacuum energy should be
determined by the \bd-branes in the throat. To this purpose, it
must be required that the initial energy $\Delta p T_3( r_*/ R
)^4$ of \bd-brane entering into the throat in unite of Hubble time
should far less than $2p\beta^4 T_3$. Thus $\Delta p (r_*/R)^4\ll
p\beta^4 $ or equally $\epsilon (r_*/R)^4\ll \beta^4 $ need to be
satisfied, which implies that a mild warping of the throat or a
large value of $p$ should be considered, which may be actually
obtained easily. For instance, we may take ${\cal K}\sim {\cal M}
\gg p \gtrsim \Delta p /\beta^4$. When entering the throat, the
\bd-branes can feel a net radial force proportional to the 5-form
flux, which will attract them to the tip at $r = r_0$. This force
is a sum of gravitation and 5-form contributions. Note that For
the D3-branes, both terms cancel. When the brane arrives at
$r=r_0$, the interaction between \bd-branes will be involved.
However, here for a simplified discussion, we only study its net
effect, i.e. the energy density driving the inflation is added
with $p$ units. In addition, we also assume here that all the
moduli have been fixed, as has been done in Ref. \cite{KKLT}.




The primordial perturbation can be generated during this evolution
induced by the increase of number of \bd-branes along $p$
direction\footnote{The reason is there can be a slicing of uniform
energy density orthogonal to the $p$ direction, which sets $
\delta\rho-\rho^{\prime}\delta \eta =0$, and thus $\delta\eta=
\delta\rho/ \rho^{\prime}$, where the prime denotes the derivative
for the conformal time $\eta$ and $\rho$ is the energy density,
thus the change of energy density along $p$ direction can induce
the comoving curvature perturbation $\zeta$ along the change
direction as $\zeta=h{\delta\rho\over {\dot \rho}}-\Phi$, where
$\Phi$ is the Bardeen potential.}. To calculate it simply but
effectively, we adopt the work hypothesis in which the null energy
condition violating evolution behaves like the phantom energy,
whose simplest implementation is a scalar field with the reverse
sign in its kinetic term.
Thus to model the evolution required, we introduce a phantom field
$\varphi$ here. Instead of the rolling down of normal scalar field
along the potential, the phantom field will be driven up along its
potential, as has been analytically and numerically shown in e.g.
Refs. \cite{GPZ0304, ST}, and at late time for some potentials,
analogous to the slow roll regime of normal scalar field, the
phantom field will enter into a ``slow climb" regime, where ${\dot
\varphi}^2\ll {\cal V}(\varphi)$. Thus we can have $h^2\simeq
{\cal V}/3$, where ${\cal V}$ is the potential of phantom field,
which can be distinguished from $V({\psi})$. In this case, if we
approximately regard $|\epsilon |$ as constant, which here means
that $\Delta p/p$ is nearly unchanged, we can have \cite{PZ, PZ1}
\be {\cal V}(\varphi)=2p_{i}\beta^4 T_3 \exp{\left(-\sqrt{\Delta
p\over p} \varphi\right)}, \label{calv}\ee where the prefactor is
determined by Eq.(\ref{h}), and $p_{i}$ is the value of $p$ at the
beginning time of phantom inflation. In Eq.(\ref{calv}),
$-\varphi$ direction corresponds to the increasing direction of
$p$ in Fig.1.
In the
slow climb approximation $|\epsilon |\simeq {({\cal
V^{\prime\prime}}/{\cal V})^2\over 2}$ \cite{PZ1}, where the prime
is the derivative with respect to $\varphi$, thus combining it
with Eq.(\ref{calv}), we have $|\epsilon|\simeq {{\Delta p}\over
2p}$, which is consistent with the requirement of our model. Thus
in this sense the upclimbing of energy density induced by the
increase of $p$ may be depicted well by the evolution of a phantom
field.

The calculations in this case have been done in Ref. \cite{PZ,
PZ1}. It can be noted that here Fig.1 is nearly same as Fig.1 in
Ref. \cite{PZ1} with a replacement $ \psi$ with $\sigma$, in which
to have an exit from the phantom inflation, a normal scalar field
$\sigma$ is introduced to implement a water fall reheating, like
that occurring in the hybrid inflation. Here we will reillustrate
the calculations in Ref. \cite{PZ1}, however, renew in a way
independent of the phantom potential, which will be helpful for
the analysis of spectral index tilt here. We will work in the
longitudinal gauge. In the momentum space, the equation of motion
of gauge invariant variable $u_k$, which is related to the Bardeen
potential $\Phi$ by $u_k\equiv a \Phi_k/\varphi^\prime $, is given
by \be u_k^{\prime\prime} +\left(k^2 - {\beta(\eta)\over
\eta^2}\right) u_k =0 \label{uk}\ee where the prime is the
derivative with respect to the conformal time $\eta$ and $
\beta(\eta)\simeq \epsilon -\left({d\ln{|\epsilon |}\over d{\cal
N}}\right)/2$, which may be obtained by taking $|\epsilon |\simeq
0$ in Eq.(21) of Ref. \cite{PZ}, where the higher order terms like
$\left({d\ln{|\epsilon|}\over d{\cal N}}\right)^2$ and
${d^2\ln{|\epsilon|}\over d{\cal N}^2}$ have been neglected since
for inflation ${d\ln{|\epsilon |}\over d{\cal N}}\ll 1$, and
${\cal N}$ is the efolding number of mode with some scale $\sim
1/k$ which leaves the horizon before the end of phantom inflation
and may be defined as \be {\cal N}=\ln{\left({a_eh_e\over
ah}\right)},\ee where the subscript 'e' denotes the quantity
evaluated at the end of the phantom inflation. Here since $h$ is
nearly constant, we can have ${\cal N}\simeq \ln{({a_e\over a})}$.
Note that for ${d\ln{|\epsilon |}\over
d{\cal N}}\ll 1$, $\beta$ is near constant for all interesting
modes $k$. Thus Eq.(\ref{uk}) is a Bessel equation and its general
solutions are the Bessels functions with the order $v=\sqrt{\beta
+{1\over 4}}\simeq {1\over 2}+\beta$, since $|\beta|\ll 1$.

In the regime $k\eta \rightarrow \infty $, the mode $u_k$ are very
deep in the horizon. Thus Eq.(\ref{uk}) can be reduced to the
equation of a simple harmonic oscillator, in which $ u_k \sim
{e^{-ik\eta} /(2k)^{3/2}}$ can be taken as the initial condition.
In the regime $k\eta \rightarrow 0$, the mode $u_k$ are far out
the horizon, and become unstable and grows. In long wave limit,
the expansion of the Bessel functions to the leading term of $k$
gives \be k^{3/2}u_k\simeq  {\sqrt{\pi}\over 2^{3/2}\sin{(\pi
v)}\Gamma(1-v)}\left({-k\eta\over 2}\right)^{{1\over 2}-v}.
\label{kuk}\ee Thus we can have $k^3|u_k|^2 \sim k^{1-2v}\sim
k^{-2\beta}$, which gives the spectral index of $\Phi$ \be
n_{\Phi}-1 \simeq -2\epsilon +{d\ln{|\epsilon |}\over d{\cal N}},
\label{ns}\ee This spectrum is actually given by the constant mode
of Bardeen potential $\Phi$ \cite{PZ2}, which may easily be
inherited by the constant mode after the reheating, and thus the
curvature perturbation $\zeta$ in comoving supersurface will have
same spectrum as $\Phi$, i.e. $n_{\zeta}=n_{\Phi}$.
When $|\epsilon |\ll 1$, we see that the spectrum is nearly
scale invariant.

The tilt of spectral index is interesting for the observations.
When $\epsilon$ is constant, the spectrum is slightly blue, since
${d\ln{|\epsilon|}\over d{\cal N}}=0$ and $\epsilon \lesssim 0$,
which looks like not favored by the observation \cite{WMAP}.
However, whether $\epsilon$ is constant is actually dependent of
the rate that the \bd-branes are sent into the throat. When this
rate is proximately constant, we will have $\epsilon$ constant, or
$\epsilon$ will be time dependent. Thus in principle the spectrum
may be blue or red, dependent of physical details of \bd-branes
motion in the compactification bulk. To make the spectrum become
slightly red, ${d\ln{|\epsilon |}\over d{\cal N}}$ is negative and
larger than $2|\epsilon| $ is required, which means that
$|\epsilon |$ must be increased with the decreasing of ${\cal N}$.
Thus it require that in per Hubble time the number $\Delta p$ of
\bd-branes entering the throat should be more and more with the
time. We show this point by a detailed example, which is plotted
in Fig.2. This result indicates that under certain condition the
spectrum may fit the observations well \cite{WMAP}. Note that the
spectral index is to large extent determined by the random motion
of \bd-branes, thus to match the observation, it seems that some
fine tuning for the motion of \bd-branes must be required,
however, in turn, if the primordial perturbation is actually
generated by such an inflation model, in some sense the spectral
index observed will encode the ordered motion of \bd-branes.

\begin{figure}[t]
\begin{center}
\includegraphics[width=9cm]{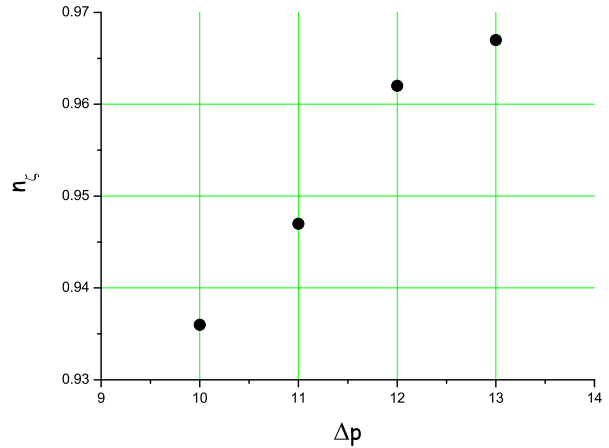}
\caption{The figure of $n_\zeta$ with respect to $\Delta p$ in per
Hubble time, which is plotted by following calculations. For per
Hubble time, we have $\Delta {\cal N} \simeq -1$, Eq.(\ref{ns})
can be written as $n_{\zeta}-1\simeq 2|\epsilon |- {\Delta
|\epsilon| \over |\epsilon|}$, since $\epsilon\lesssim 0$. We take
${\cal M}\sim 10^4$ and $p\simeq 500$ as example, as has been
mentioned in the text. In this case, in two subsequent Hubble
times, we take the numbers of \bd-branes entering the throat e.g.
be 10, 11. For $\Delta p= 10$, we have $|\epsilon_{10}| \simeq
{\Delta p\over 2p} \simeq 0.01$. In the subsequent Hubble time, if
$\Delta p= 11$, we will have $|\epsilon_{11}| \simeq 0.0108$, thus
$\Delta|\epsilon|\simeq 0.0008$, which gives $n_{\zeta}\simeq
1-0.053$. 
}
\end{center}
\end{figure}

The amplitude of curvature perturbation $\zeta$ may be given by
\ba {\cal P}_{\zeta} & = & {k^{3}\over 2\pi^2}|\zeta_k| \simeq
{1\over 2\pi^2|\epsilon|^2}\cdot {1\over 2^3}
\left|{\varphi^{\prime}\over a }\right|^2 \nonumber\\ & \simeq &
{1\over 2|\epsilon|}\left({h\over 2\pi}\right)^2, \label{phik}\ea
where ${\dot h}={\dot \varphi}^2/2$ and Eq.(\ref{kuk}) have been
used, in addition we also use $\zeta \simeq \Phi/\epsilon$ since
$|\epsilon|\ll 1$. This may be seen by using the relation between
$\zeta$ and $\Phi$, which is $ \zeta\equiv \Phi+{1\over
\epsilon}\left({a\Phi^\prime\over a^{\prime}}+\Phi\right)$. Thus
when $|\epsilon|\ll 1$, it naturally leads to $\zeta\simeq
\Phi/\epsilon$. When $|\epsilon|\gg 1$, we have $\zeta\simeq
\Phi$, which has been used in the calculations of primordial
perturbations generated during a slow expansion with $\epsilon\ll
-1$ \cite{PZ}. Eq.(\ref{phik}) is the same as that of usual
inflation models.

The spectrum of tensor perturbation can be calculated in a similar
way \cite{PZ1, Piao0601}. The amplitude of its spectrum is given
by \be {\cal P}_{\rm T} \cong {k^3\over 2\pi^2}\cdot
4\cdot\left|{\sqrt{2}v_k\over a}\right|^2 \simeq 8\cdot
\left({h\over 2\pi}\right)^2, \label{ptt}\ee where the gauge
invariant variable $v_k$ is related to the tensor perturbation
$h_k$ by $v_k = a h_k/\sqrt{2}$. Thus the ratio of tensor to
scalar perturbation is $r ={\cal P}_{\rm T}/{\cal P}_{\zeta}\simeq
16 |\epsilon|$. Here in unit of Hubble time $|\epsilon | \simeq
{\Delta p\over 2p}$, thus $r\simeq {8\Delta p\over p}$, which can
be large, whose detailed value depends on the change of $\Delta
p/p$. In general in this model $p\sim 10^2$ to $10^3$ and
$1\lesssim\Delta p \lesssim 10$ or several $10$. $\Delta p \gtrsim
1$ is because in probability in unit of Hubble time there should
be at least one \bd-brane entering into the throat. This actually
sets a lower limit for $r$. In our example, $p\simeq 500$, thus
$r\gtrsim 8/500\simeq 0.016$, which suggests that if $r$ is enough
small, for example $r<0.01$, this model will be ruled out. $\Delta
p \lesssim 10$ or several $10$ is to make $|\epsilon|\ll 1$ valid
in all time. In our example, $r\simeq 8\times 10/500 =0.16$, which
is consistent with recent observation \cite{WMAP5}, in which
$r<0.20$. Thus in principle our model predicts a mild range of
$r$, i.e. $0.01\lesssim r \lesssim 0.2$, dependent of the details
of model, which may be tested in future observations.

The spectral index of tensor perturbation is nearly scale
invariant with the tilt equal to $-2\epsilon$ \cite{PZ1}. Here
note that $\epsilon\lesssim 0$, thus the tensor spectrum is
slightly blue, which is a distinct feature, since for the usual
inflation models, the tensor spectrum is generally slightly red.
However, this feature can be consistent with the observations
\cite{FRM}.

In principle, it seems that we also may introduce other fields
with any kinetic term or fluids to describe this null energy
condition violating evolution, as long as their energy density is
increased with the time. However, note that for such fields or
fluids, generally the sound speed $c_s^2$ is not exact 1 in their
perturbation equation. This means in this case the causal
structure of background and perturbation evolution has been
altered, which actually is not expected to occur in our model,
since here there seems not any motion or evolution leading to such
a change of the causal structure. Further, there might exist some
fluid which violates the null energy condition and in the meantime
whose parameter $w$ of state equation and $c_s^2$ are independent
each other, and then $c_s^2$ can be set as 1. However, in this
case we will obtain same Eq.(\ref{uk}), where $u_k\equiv
\Phi_k/\sqrt{\rho+p} $, and thus same Eqs.(\ref{phik}) and
(\ref{ptt}). This means that this case is actually equal to that
of the simple phantom field. Thus using the perturbation equation
derived for the phantom field can be reasonable for our model.


In summary, we note that in a class of warped compactifications
with the brane/flux annihilation, there may be a flat direction
along the number $p$ of \bd-branes locating at the tip of KS
throat, which may drive a period of phantom inflation, since in
the four dimension effective description the upclimbing of energy
density along $p$ direction corresponds to a null energy condition
violating evolution. The spectrum of adiabatic perturbation can be
nearly scale invariant with a mild tilt dependent of the change of
\bd-branes number entering into the throat.
This work seems slightly idealistic, however, it may be a start,
with which it may be expected that one could envisage more
examples in string theory, in which the four dimension effective
descriptions may share some similar characters as the phantom
inflation. We have neglected some details involving the motion of
the \bd-branes and the interaction between them, which maybe able
to lead to some different features of primordial spectra. This
will be left in the future works.


\textbf{Acknowledgments} The author thanks Y.F. Cai for comments.
This work was motivated during the workshop in KITPC and is
supported by KITPC, and also in part by NNSFC under Grant No:
10775180, in part by the Scientific Research Fund of
GUCAS(NO.055101BM03), as well as in part by CAS under Grant No:
KJCX3-SYW-N2.

\end{document}